\documentclass[11pt,twoside]{article}

\usepackage{asp2004}
\usepackage{epsf}
\usepackage{psfig}
\usepackage{lscape}
\usepackage{graphicx}

\begin{document}
\title{The Spectropolarimetric Evolution of V838 Monocerotis}   
\author{John P. Wisniewski}   
\affil{USRA/NASA Goddard Space Flight Center, Code 667, Greenbelt, MD 20771 USA; jwisnie@milkyway.gsfc.nasa.gov}

\begin{abstract} 

I review photo-polarimetric and spectropolarimetric observations of V838 Mon, 
which revealed that it had an asymmetrical inner circumstellar envelope 
following its 2nd photometric outburst.  
Electron scattering, modified by pre- or post-scattering H absorption, is the polarizing mechanism 
in V838 Mon's envelope.  The simplest geometry implied by these observations is that of a 
spheroidal shell, flattened by at least 10\% and having a projected position angle on the sky of
$\sim$37$^{\circ}$.  Analysis of V838 Mon's polarized flux reveals that this electron scattering shell 
lies interior to the envelope region in which H$\alpha$ and Ca II triplet emission originates.  To 
date, none of the theoretical models proposed for V838 Mon have
demonstrated that they can reproduce the evolution of V838 Mon's inner circumstellar environment,
as probed by spectropolarimetry.

\end{abstract}


\section{Diagnostic Capabilities of Polarimetry}   

Linear polarimetry can provide powerful diagnostic information regarding the 
geometry of unresolved astrophysical environments.  Numerous literature resources 
\citep{nor92, bjo00} eloquently discuss these diagnostic capabilities; for the 
purpose of this review I will simply summarize several fundamental principles.  

The observed intrinsic 
polarization of unresolved sources is simply the net integrated polarization 
of the system.  The density, geometrical distribution, and scattering properties 
of scatterers in a system are several factors
which will influence the strength of the observed intrinsic polarization; systems
which either lack an extended envelope of material or are 
characterized by a symmetrical envelope will exhibit zero net intrinsic 
linear polarization.  Non-uniform illumination of an extended envelope, by sources 
such as star-spots and/or binary companions, may also produce a net intrinsic 
polarization.  

Several factors may influence the wavelength dependence of observed intrinsic linear 
polarization, as discussed by \citet{nor92} and \citet{bjo00}.  These 
factors include: a)
the scattering process (i.e. Thompson versus Mie scattering); b) 
the nature of the illuminating source;
c) the dilution of polarized light by the presence of additional unpolarized (i.e. direct) 
light; and d) 
the preferential absorption of more scattered light than direct (unpolarized) light.

One is typically is unable to directly measure the intrinsic 
polarization of astrophysical sources, as the actual observed polarization is comprised 
of interstellar (time independent) and intrinsic (possibly time dependent) components.  
Identifying and removing interstellar polarization (ISP) from data is a critical, 
non-trivial exercise; however, several techniques have proven to be successful in this regard. 
The field star technique \citep{mcl79} is one method; 
successful implementation requires one to identify a suitable number of field stars which are
a) intrinsically unpolarized; b) located at a similar distance as the target of interest; 
and c) located a small angular distance from the target of interest.  If one assumes emission 
lines in the target of interest are intrinsically unpolarized \citep{har68}, measuring the 
polarization in these lines can yield estimates of the ISP, although \citet{qui97} have shown 
that this assumption is not always valid.  Finally, the wavelength dependence of ISP is 
known to follow the empirical Serkowski law \citep{ser75}; hence, measuring the wavelength 
dependence of the total observed polarization will, in certain cases, allow one to 
accurately parameterize the ISP.

\section{Observational Datasets}   

Four groups have reported polarimetric observations of V838 Mon \citep{wi03a, 
wi03b,des04,rus05}.  While these studies generally yield similar results regarding the 
temporal polarimetric evolution of V838 Mon and the fundamental mechanism responsible
for this polarization, there is some disagreement amongst the datasets.   
One likely source of these differences is the level of instrumental polarization which
characterize these datasets and act as a source of systematic errors.
The absolute accuracy of the \citet{wi03a}, \citet{wi03b} and \citet{rus05} 
datasets are 0.025\% and 1 degree at V, $<$ 0.05\% and 1 degree, and 0.08\% and $\sim$1 
degree at V respectively.  The instrumental 
polarization of the Asiago data of \citet{des04} is quoted as 0.2\% and 2 degrees, 
although \citet{for06} quote an instrumental polarization of $<$0.4\% for this instrument.
The instrumental polarization of the Crimean data of \citet{des04} is quoted as ranging 
from 2.4\% (U) to 0.5\% (I), with a scatter of 0.1\% (U) and 0.02\% (other bands).  In 
addition, \citet{des04} note an \textbf{additional 6 degree position angle offset} 
between their Asiago and Crimean datasets.

Interestingly, the polarization magnitude and position angle of the dataset of 
\citet{des04} is different by $\sim$0.2\% and $\sim$6 degrees from other published 
polarimetry \citep{wi03a,rus05}, corresponding almost exactly to the additional 
level of instrumental polarization in their data.  It is possible that the differences 
between these datasets are related to systematic errors present in the dataset of 
\citet{des04}; the specific intrinsic polarization values reported by
\citet{des04} might be suspect.  

\section{Spectropolarimetric Evolution: Timeline}   

\begin{figure}
\begin{center}
\includegraphics[width=9cm]{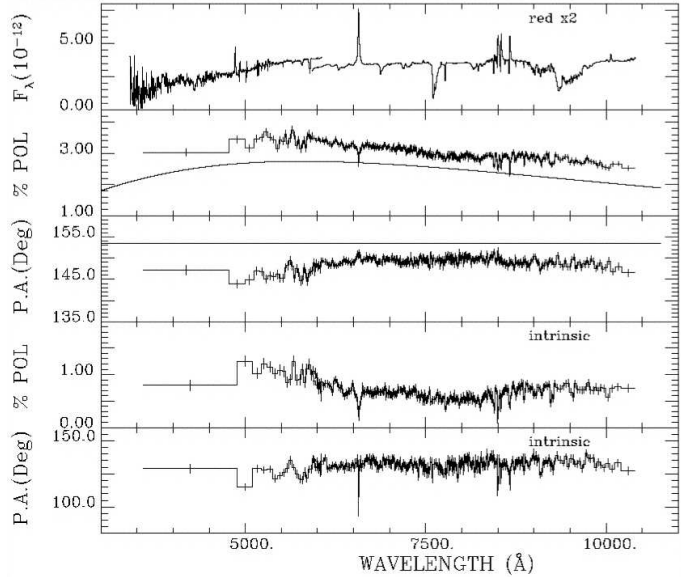}
\caption{The flux (top panel), total polarization (panels 2 and 3), and intrinsic 
polarization (bottom 2 panels) of V838 Mon on 8 Feb 2002 (adopted from \citealt{wi03a}).  
The solid line in the total polarization plots (panels 2 and 3) represent the best-fit 
interstellar polarization component of \citet{wi03a}.
 \label{intrinpol}}
\end{center}
\end{figure}

\begin{figure}
\begin{center}
\includegraphics[width=9cm]{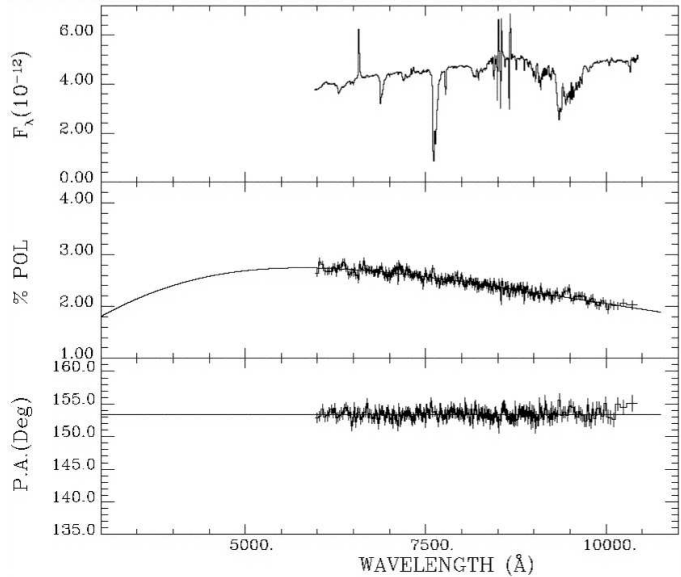}
\caption{The flux (top panel) and total polarization (bottom 2 panels) of V838 Mon on 
13 Feb 2002, along with the best-fit Serkowski-law interstellar polarization estimate 
(solid line), as adopted from \citet{wi03a}.  
\label{ismpol}}
\end{center}
\end{figure}

V838 Mon exhibited clear evidence of variability in its total observed polarization 
in early 2002 \citep{wi03a,des04}.  Below I summarize the notable events in 
V838 Mon's spectropolarimetric behavior: \begin{itemize}

\item 10 and 11 Jan 2002; The integrated V-band polarization is measured to be within 1-$\sigma$ of 
the (later to be known) interstellar polarization value \citep{des04}; it is 
unclear whether any intrinsic component existed at this early epoch.

\item 4 Feb 2002; The integrated V-band polarization measured by \citet{des04} is 
well above the (later to be known) interstellar polarization value; this is the first 
clear evidence of the presence of an intrinsic polarization component.

\item 8 Feb 2002; The first spectropolarimetric observations of V838 Mon confirm the 
presence of an intrinsic polarization component.  The magnitude of the observed 
total polarization is clearly higher than (later to be known) interstellar values  
(see panel b of Figure \ref{intrinpol}).  \citet{wi03a} find the integrated  
R-band polarization to be 3.2\% at a position angle (PA) of 149$^{\circ}$. 
Observed line depolarization effects at H$\alpha$ and the Ca II triplet are also 
indicative of an intrinsic component.  Subsequent spectropolarimetric observations 
on 11 Feb 2002 \citep{des04} confirm these results.

\item 13 Feb 2002; Spectropolarimetry by \citet{wi03a} suggest that V838 Mon's intrinsic 
polarization component has disappeared.  These data clearly 
exhibit a Serkowski-law \citep{ser75} wavelength dependence, the best fit of which is 
shown as a solid line in Figure \ref{ismpol}.  \citet{wi03a} measure the integrated R-band 
polarization to be 2.667\% at a PA of 153$^{\circ}$.  Line polarization effects 
have disappeared (Figure \ref{ismpol}), 
supporting the suggestion that the polarization is purely interstellar in origin.  

\item 14 Feb 2002; Spectropolarimetry by \citet{rus05} confirm that V838 Mon's polarization
appears to be purely interstellar in origin at this time.  A second observation by this 
group on 7 Mar 2002 finds a similar result, although the authors do not rule out the 
possibility that a very small intrinsic component is still present.

\item 15 Feb - 20 Mar 2002; Polarimetric observations by \citet{des04} detect no clear 
evidence of an intrinsic polarization component, rather the data appear to be characterized 
purely by an interstellar polarization component. 

\item 22-24 Oct 2002; Photo-polarimetric observations by \citet{wi03b} suggest the renewed 
presence of an intrinsic polarization component.  The position angle of this intrinsic polarization
component is oriented 90$^{\circ}$ from that observed on 8 Feb 2002, suggesting either a 
fundamental change in the illumination of the system's scatterers or a fundamental change in 
the geometrical distribution of these scatterers. \end{itemize}

\section{Interstellar Polarization} 

\citet{wi03a} fit a modified Serkowski-law \citep{ser75,wil82} to their 
13 Feb 2002 data (Figure \ref{ismpol}), yielding the ISP parameters: 
P$_{max}$ = 2.75 $\pm$ 0.1\%, $\lambda_{max}$ = 5790 $\pm$ 37\AA, 
PA = 153.4 $\pm$ 0.12$^{\circ}$, K = 0.971, and dPA = 0.  Field star 
ISP measurements \citep{wi03b} and later epoch  
spectropolarimetry \citep{rus05} support this ISP determination.
While it has been suggested in the literature \citep{des04} that the line 
depolarization effects in the \citet{wi03a} dataset do not visually fit their ISP 
determination, careful analysis of these data reveals this discrepancy 
is less than a 1-$\sigma$ effect.  Thus the line depolarization effects seen in 
Figure \ref{intrinpol} do support the ISP determination of \citet{wi03a}.  
\citet{des04} suggest a modestly different ISP; however, as discussed in 
Section 2, systematic errors in their dataset are likely to be the source of  
this discrepancy. 

\subsection{Light Echo Material: Interstellar vs. Circumstellar}

Interstellar polarization is produced by
the dichroic absorption of starlight by aligned interstellar 
dust grains.  During this conference, it was suggested that the 
large interstellar polarization component associated with V838 Mon supports the notion that 
its light echo material is interstellar in origin.  However the observed interstellar polarization 
is merely a superposition of all absorption events which occur along a line of sight; thus,  
V838 Mon's interstellar polarization provides neither evidence for an interstellar nor a 
circumstellar origin of its light echo material.

\section{Intrinsic Polarization}

\subsection{Polarizing Mechanism}

The wavelength dependence of V838 Mon's intrinsic polarization (bottom 2 panels of 
Figure \ref{intrinpol}) exhibits several features: a) the polarization magnitude is 
nearly constant with wavelength, but does seem to slowly rise at blue wavelengths; b) there 
is 1-$\sigma$ evidence of a polarization jump at the Paschen limit \citep{wi03a}; and c) 
there is little evidence of intrinsic polarization in any of the strong emission lines.  
These unique spectropolarimetric signatures are similar to those observed for classical
Be stars \citep{woo96,woo97}, leading both 
\citet{wi03a} and \citet{des04} to suggest that the mechanism polarizing V838 Mon's  
environment is electron scattering, modified by pre- or post-scattering absorption by 
hydrogen.  The behavior of the electron scattering wings in the H$\alpha$ line profiles of 
V838 Mon, which were present while V838 Mon exhibited an intrinsic polarization component and 
disappeared 1 day after the disappearance of this intrinsic polarization \citep{wi03a}, further 
supports this interpretation.  Other polarizing mechanisms such as Rayleigh scattering or 
scattering by dust grains would have
produced a significantly different wavelength-dependent polarization signature than that 
observed.

\subsection{Geometry of the Scatterers}

The presence of an intrinsic polarization component implies that V838 Mon's circumstellar envelope 
deviates from spherical symmetry.  \citet{wi03a} note that one of the possible geometries of this 
envelope is that of a flattened spheroidal shell.  \citet{bro77} and \citet{cas87} developed tools 
to estimate the polarization produced by optically-thin electron scattering envelopes.   
\citet{bjo94} used this approach, introducing several assumptions regarding the structure and optical 
depth of the envelope, to obtain a rough estimate of the amount of flattening in the envelope of Nova Cygni 1992.
To summarize these calculations (K. Bjorkman 2006, personal communication), the flattening required 
to produce a polarization p is \begin{equation} \frac{2(a-b)}{a} = 
\frac{(20*p)}{< \tau > sin^{2}i} \end{equation}   We used this technique to obtain a crude estimate of 
V838 Mon's environment; 
\textbf{the \textit{minimum} amount of flattening of V838 Mon's spheroidal shell on 8 February 2002 
is 10\%}, given its intrinsic V-band polarization (0.98\%, \citealt{wi03a}). 
The intrinsic polarization PA on 8 Feb was 127$^{\circ}$ \citep{wi03a}, indicating 
that the projected PA of this flattened shell on the sky was 37$^{\circ}$.  Interestingly, this
position angle corresponds to the position angle derived from interferometric observations of 
V838 Mon, specifically the ``binary'' model of these data \citep{lan05}.

\subsection{Location of the Polarizing Scatterers}

\begin{figure}
\begin{center}
\includegraphics[width=9cm]{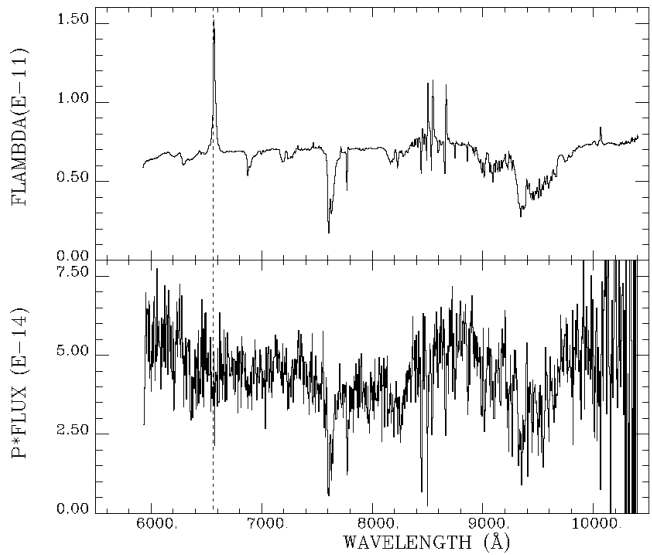}
\caption{The flux (top panel) and polarized flux (bottom panel) of HPOL spectropolarimetry 
from 8 Feb 2002.  Emission features present in the total flux but \textit{absent} in the
polarized flux spectrum, such as H$\alpha$ and the Ca II triplet, must be formed in regions
external to the flattened spheroidal shell of electrons producing the observed polarization. \label{pflux}}
\end{center}
\end{figure}

Having established that electron scattering, modified by pre- or 
post-scattering absorption by hydrogen, is the mechanism responsible for V838 Mon's intrinsic 
polarization, we now consider the location of this material. 
The (intrinsic) polarized flux spectrum 
is merely the spectrum of the illuminating source as seen by the scatterer.  Hence, 
any emission feature which is seen in the spectrum but not in the polarized flux must be external to 
the scatterer (K. Nordsieck 2006, personal communication).  As seen in Figure 
\ref{pflux}, H$\alpha$ and the Ca II triplet are clearly in emission 
in V838 Mon's spectrum on 8 Feb 2002 (top panel), while these features are \textbf{absent} from 
the polarized flux (bottom panel).  Thus, in early Feb 2002, the suggested flattened spheroidal shell of electrons 
responsible for producing V838 Mon's polarization must lie \textbf{interior} to the extended 
envelope producing the various emission features observed in V838 Mon's spectrum.

\section{Intrinsic Polarization at Later Epochs} 

As noted in Section 3, \citet{wi03b} reported photo-polarimetric observations taken on 22-24 Oct 2002 
which strongly suggested that V838 Mon had redeveloped an intrinsic polarization component.  Interestingly, 
the intrinsic PA of these data was oriented 90$^{\circ}$ from that observed on 8 Feb 2002.  \citet{wi03b} 
suggest that this renewed intrinsic polarization might arise from (1) a new source of asymmetrical 
scatterers in V838 Mon's envelope; (2) fundamental changes in the opacity of V838 Mon's envelope; (3) 
a change in the illumination source, possibly related to the emergence of the B-type binary in 
V838 Mon's spectrum during this time-period \citep{des02,mun05}; or (4) a combination of 
these scenarios.  Continued photo-polarimetric or spectropolarimetric monitoring may elucidate the 
source of this intrinsic polarization. 

\section{Summary}

V838 Mon exhibited clear evidence of an intrinsic polarization component beginning at least on 
4 February 2002; evidence of this intrinsic component disappeared by 13 February 2002.  The 
wavelength dependence of this intrinsic polarization suggests that electron scattering, 
modified by pre- or post-scattering absorption by hydrogen, was the polarizing mechanism.  The 
presence of an intrinsic component implies that V838 Mon's circumstellar envelope was asymmetrical; 
one possibly geometry of this envelope is a spheroidal shell flattened by at least 10\%.  From 
V838 Mon's polarized flux, we know that this shell is located interior to the region of V838 Mon's 
envelope responsible for producing emission features at H$\alpha$ and the Ca II triplet.  Current 
theoretical efforts to identify the mechanism responsible for V838 Mon's outburst have 
primarily focussed on explaining its photometric evolution.  To date, none of these models have 
demonstrated that they can reproduce the evolution of V838 Mon's inner circumstellar environment,
as probed by spectropolarimetry.

\acknowledgements 
I thank the present and past staffs of the Pine Bluff Observatory, Ritter Observatory, and CTIO  
who contributed to various aspects of the work presented here.  I also thank Karen Bjorkman 
for her assistance in the interpretation of these datasets.  Partial financial support to 
attend this conference was provided by the conference organizers and by an 
international travel grant from the AAS.


\question{L. Bernstein} How non-spherical does the scattering shell need to be to be consistent with 
the polarization data?
\answer{J. Wisniewski} A crude estimate is that the shell is flattened by $\sim$10\%, as detailed  
in Section 5.2 of this write-up.

\end{document}